\documentclass[conference]{IEEEtran}
\IEEEoverridecommandlockouts
\usepackage{cite}
\usepackage{amsmath,amssymb,amsfonts}
\usepackage{algorithmic}
\usepackage{graphicx}
\usepackage{textcomp}
\usepackage{xcolor}
\usepackage{hyperref}
\usepackage{graphicx}
\usepackage{booktabs}
\usepackage{csquotes}
\graphicspath{ {./} }
\def\BibTeX{{\rm B\kern-.05em{\sc i\kern-.025em b}\kern-.08em
    T\kern-.1667em\lower.7ex\hbox{E}\kern-.125emX}}

\usepackage{etoolbox}
\makeatletter
\patchcmd{\@makecaption}
  {\scshape}
  {}
  {}
  {}
\makeatother

\usepackage[margin=2cm]{geometry}
\usepackage[most]{tcolorbox}

\newtcolorbox{mytextbox}[1][]{%
  sharp corners,
  enhanced,
  colback=white,
  height=2.2cm,
  attach title to upper,
  #1
}

\begin{document}

%\title{A Radically Experimental Approach to Model Design and Refinement in Distributed Systems}
\title{A Deployment-First Methodology to Mechanism Design and Refinement in Distributed Systems}

\author{Martijn de Vos, Georgy Ishmaev, Johan Pouwelse, Stefanie Roos\\Delft University of Technology, The Netherlands}

\maketitle

\begin{abstract}
Catalyzed by the popularity of blockchain technology, there has recently been a renewed interest in the design, implementation and evaluation of decentralized systems.
Most of these systems are intended to be deployed at scale and in heterogeneous environments with real users and unpredictable workloads.
Nevertheless, most research in this field evaluates such systems in controlled environments that poorly reflect the complex conditions of real-world environments.
In this work, we argue that deployment is crucial to understanding decentralized mechanisms in a real-world environment and an enabler to building more robust and sustainable systems.
We highlight the merits of deployment by comparing this approach with other experimental setups and show how our lab applied a \emph{deployment-first methodology}.
We then outline how we use Tribler, our peer-to-peer file-sharing application, to deploy and monitor decentralized mechanisms at scale.
We illustrate the application of our methodology by describing a deployment trial in experimental tokenomics. 
Finally, we summarize four lessons learned from multiple deployment trials where we applied our methodology.

\end{abstract}


\section{Introduction}

The scale and complexity of distributed systems have increased tremendously since the field's inception.
We live in an era of complex ultra-large-scale networks characterized by the number of participating nodes and by high heterogeneity, flexibility, non-trivial social dependencies, and emergent properties~\cite{vespignani_predicting_2009,Gabriel_2006,blair_complex_2018}.
Ensuring the proper functioning, deployment, monitoring and maintenance of such systems requires system designers to obtain insights into their performance and correctness. Given the scale and unpredictability of such systems, obtaining engineering insights presents unique challenges for researchers and developers.

Decentralized blockchain applications embody the characteristics of complex ultra-large-scale networks~\cite{tessone_stochastic_2021, blackburn_cooperation_2022}.
At the same time, the costs of failures in these applications are very high due to built-in financial mechanisms~\cite{zhou_sok_2022}.
Though realistic experimental setups such as testnets aim to address these challenges to a degree, properly testing and evaluating such systems remains a challenging endeavour~\cite{fisman_formal_2022}.

The mainstream paradigm in distributed systems research is a top-down design that focuses on predicting performance, failures and limitations through experimentation in controlled environments.
These experiments usually are carried out as simulations or emulations informing researchers' design choices~\cite{buchert_survey_2015}. 
An empirical study of failures can provide invaluable insights into different types of distributed systems~\cite{gupta_failures_2017,garraghan_emergent_2018,wang_what_2017}.
Detection of events that make a system fail to operate according to its specifications - detection of failures - is often a critical task in distributed systems given complex interdependencies\cite{javadi_failure_2013}.

However, empirical experimentation, guided by workloads extracted from a deployed system, is uncommon in academic research~\cite{buchert_survey_2015}. Mature research methodologies with an emphasis on deployment are still missing even in empirically-driven fields of distributed systems research, for example, blockchain applications~\cite{marijan_blockchain_2022}. 

We address this gap by presenting our deployment-first methodology for designing and evaluating decentralized systems.\footnote{The scope of this paper is primarily limited to decentralized systems. However, some of these insights could be relevant to a wider field of distributed systems.} 
We specifically focus on findings that can be obtained from the study of failures after the deployment.
Our methodology is based on nearly two decades of experience developing the Tribler software~\cite{pouwelse2008tribler,zeilemaker2011tribler,tribler_source_code}, serving as infrastructure for deploying and evaluating decentralized mechanisms at scale.
We demonstrate that while the deployment-first research methodology can be demanding in terms of time investment, additional insights obtained are worth this time investment. A nuanced evaluation of trade-offs between different experimental setups is also instrumental in designing research methodologies.

\begin{figure*}[!t]
\centering
\includegraphics[width=.75\linewidth]{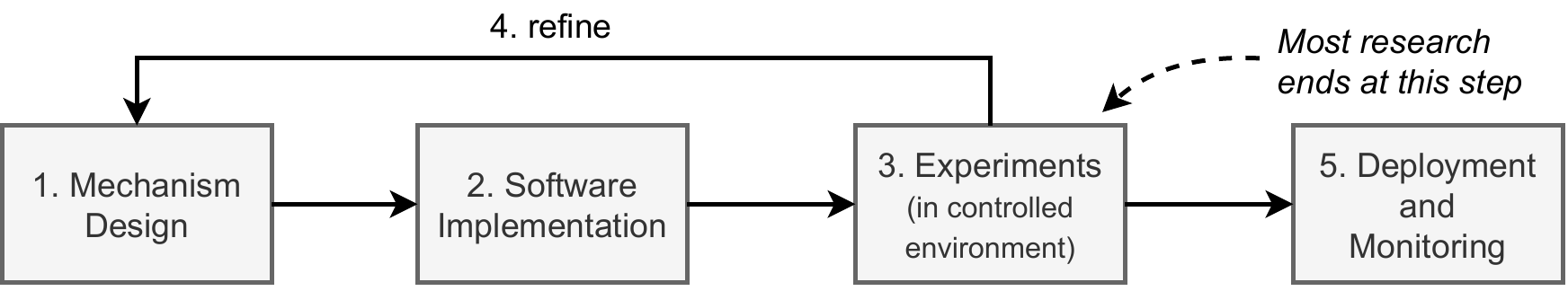}
\caption{The standard research methodology to design, build, evaluate and deploy decentralized systems.}
\label{fig:traditional_methodology}
\end{figure*}

In summary, this work makes the following contributions:

\begin{enumerate}
    \item We compare different experimental setups and highlight the merits of deployment as a critical step in the research methodology when building and evaluating distributed systems (Section~\ref{sec:experiments_distributed_systems}).
    \item We formulate our \emph{deployment-first research methodology} to evaluate decentralized mechanisms at scale (Section~\ref{sec:tribler}). We also describe how we use Tribler, our decentralized file-sharing application and research vehicle for conducting large-scale deployment trials.
    \item We illustrate how we applied our research methodology to obtain unique insights in a complex use case on tokenomics in decentralized networks (Section~\ref{sec:use_case_tokenomics}).
    \item Based on our experiences with our methodology, we summarize four lessons we learned (Section~\ref{sec:lessons_learned}).
\end{enumerate}

%The paper is structured as follows. In Section 2 we discuss how our deployment-focused approach yield unique insights compared to more traditional research cycles. We then introduce Tribler, in Section 3, our research vehicle that allows us to quickly deploy new algorithms and mechanism to thousands of active daily users. We end with a Section 4, presenting a case study that describes how we used our research methodology to obtain new insights in the complicated dynamics in the context of bandwidth accounting. We conclude with the key insights on further empirical experimentation in distributed systems. 

% P2P systems will eventually be executed at the end-user machine.
% P2P systems are highly hetrogeneous. Experiments in controlled environments are often done in homogeneous environments.

%\section{Decentralized Systems at Scale}
\section{The Status Quo of Decentralized Systems Experiments}
\label{sec:experiments_distributed_systems}
This work argues that deployment experiments are essential to build robust decentralized mechanisms.
We first describe the standard methodology to research distributed systems and then outline the merits of deployment as part of the research methodology.

\subsection{Standard Research Methodology}
Figure~\ref{fig:traditional_methodology} shows the standard research methodology to design, build, evaluate, and deploy decentralized systems.
This research methodology is based on reports in the fields' literature~\cite{Haeberlen2006FallaciesIE,gustedt_experimental_2009,buchert_survey_2015}, on discussions with other researchers, and our own experiences.
The following five steps describe this methodology:
\begin{enumerate}
    \item \textbf{Mechanism Design.} A researcher starts by designing a particular mechanism. Research with an exclusive focus on understanding theoretical models is usually limited to this step~\cite{buchert_survey_2015}.
    \item \textbf{Software Implementation.} The researcher then works on a software implementation of the designed mechanism. Assuming that the original design was well-executed, the implementation phase should not result in significant changes to the original models. As such, we left out this feedback loop from Figure~\ref{fig:traditional_methodology}.
    \item \textbf{Experiments.} With the implementation, the researcher conducts experiments in an environment controlled by the analyst, e.g., on a local computer or a compute cluster. These experiments usually aim to verify the implementation's correctness and quantify system metrics, e.g., scalability and fault tolerance.
    \item \textbf{Refining The Design Using Experimental Results.} Based on the experimental results obtained in the previous step, the researcher updates the mechanism design, updates the accompanying implementation and re-runs experiments. For instance, an experiment revealing that a particular design has low scalability (e.g., in the number of participants the system can support) might require the researcher to identify bottlenecks in the mechanism and resolve them.
    \item \textbf{Deployment and Monitoring.} The researcher can deploy the algorithm in a real-world setting after the experiments are finished. The deployed software will likely be continuously monitored to detect failures or anomalies.
\end{enumerate}

%Because of the costs of deployment this is the most usual methodology. But actual tradeoffs are more nuanced. 

Most academic research does not further test and evaluate their mechanisms using deployment and ends at step (4) \cite{buchert_survey_2015}.
This is not unexpected since local experiments usually suffice to prove the mechanism's trade-offs, correctness or performance to a scientific community.
As such, the time investment and resource costs do not justify the need for deployment.\footnote{In contrast to that, the deployment of novel system designs is a common practice in the blockchain industry\cite{marijan_blockchain_2022}. However, the quality of experimental evaluations is lacking compared to academic research, as industry whitepapers often present biased and inflated results \cite{nasrulin2022gromit}.}
We argue, however, that trade-offs and limitations associated with the usage of experimental setups to evaluate decentralized systems are more nuanced.

\begin{table*}[!tbh]
    \normalsize
	\centering % <-- important
	
	\begin{tabular}{ | l | c | c | c | c | } 
		\toprule
		\textbf{Property} & \textbf{Simulation} & \textbf{Emulation} & \textbf{Testnet} & \textbf{Real-world Deployment} \\ \hline
        Cost of experiment & Medium & Medium & Low/High & Low/High \\
        Scalability & Medium/High & Resource constrained & Resource constrained & Resource constrained \\
        Environmental realism & Low & Low/Medium & High & Very High \\
        Failures discoverability & Impossible & Low & Medium & High \\
        Reproducibility & High & Medium & Low & Low \\
        Control & High & High & Medium & Low \\
        Speed of change & Fast & Fast & Medium & Slow \\
        Debugability & High & Medium & Medium & Low \\
        \bottomrule
	\end{tabular}\vspace{0.15cm}
	\caption{A comparison between four different experimental setups: simulation, emulation, testnet, and real-world deployment.}
	\label{tab:experimental_setups_comparison}
\end{table*}

\subsection{The Merits of Deployment}
We start by comparing the trade-offs between different experimental setups used to evaluate decentralized systems.
Based on our literature research, we choose to compare the following four experimental setups: 

\begin{enumerate}
    \item \textit{A simulation} is a model of an application tested on a model of an environment;
    \item \textit{An emulation} is an application that runs in an environment where some parts of it are modelled;
    \item \textit{A testnet} is an application that is deployed on multiple machines run by researchers or volunteer testers;
    \item \textit{A real-world deployment} is an application that end users run on their machines.
\end{enumerate}

%The scale of difference between results and limits of different setups is considered in relational, not absolute values. Our heuristic applies to generic distributed systems, and absolute values are necessarily context specific.

We compare in Table~\ref{tab:experimental_setups_comparison} eight different properties of these experimental setups and briefly discuss them below.

\textbf{Costs of experiments.} It is not always feasible to represent the costs of experiments in commensurable scales. Most often, the costs of experiments can be described by the monetary costs of purchasing necessary computation resources. Both in simulation and emulation, these costs are relatively manageable.
However, in the case of a deployed system, experimentation costs can be much higher or lower, depending on the application context. Experiments on a deployed system that require changes in system parameters or functionality can cause failures and loss of users or market share~\cite{fisman_formal_2022}. However, experiments on a deployed system can have meager costs in some situations, e.g. if volunteer users provide their resources.\footnote{See \url{https://github.com/ethereum/ropsten/blob/master/revival.md}.} In testnets, if most of the resources are provided by volunteer users, the experiment costs can be low for researchers. However, attracting a sufficient amount of users for a testnet can also require some initial investments and upfront costs to get traction, as can be the case with marketing costs for blockchain testnets~\cite{ethereumgrants}.

\textbf{Scalability.} of experiments is strongly correlated with the costs of experiments. In the case of simulations, it is relatively cheap to scale up the simulation size. In emulation, the upper bound for scalability is typically limited by the available computational resources of a testbed. In the case of a testnet, the scalability is limited to the computational resources available to the researcher.
In a real-world deployment, the scale of the experiment is usually limited by the number of end-users and the resources they contribute.

\textbf{Environmental realism.} This is one of the two key features that set testnets and deployments apart from other experimental setups. Environmental realism is comprised of three different parameters: (1) client heterogeneity, e.g., differences in hardware capabilities; (2) variability in external parameters such as network conditions; (3) the effect of user behaviour.
A real-world deployment captures all these three parameters.
The key difference with the testnet is the effect of user behaviour; e.g., a testnet can be exclusive to expert users, reducing realism. User incentives in testnets are also sometimes simplified, e.g., in blockchain testnets, there usually are no financial incentives. Both in simulation and emulation, all three parameters of environmental realism are lower compared to other experimental setups. External parameters such as network conditions can be more realistic with emulation. Realistic client heterogeneity is challenging to represent realistically in an emulation conducted with heterogeneous hardware.
One observation is that environmental realism can be improved for simulation and emulation if these setups are designed with values known from measurements of deployed systems. Two key factors limit such measurements: first, the measured system should have a very similar application context; Second, it is not certain if the results of the measurements still hold as the system evolves.

There has been some work that aims to bring environmental realism to simulation or emulation setups. Sarzyniec et al. present Distem, a virtualisation platform to enable resource heterogeneity in a homogeneous compute cluster~\cite{sarzyniec2013design}.
While this is a step to make experiments more accurate, the failure model and user-generated workloads are not carried over from the real-world environment.
Recent work on scalability experiments with BFT consensus protocols proposes a simulator~\cite{berger2022does}. Understanding the \emph{realism gap} between simulators and real-world environments is a key part of this work.

\textbf{Discoverability of new failures.} This is another key property that distinguishes controlled and uncontrolled experiment setups. Certain types of failures can only be discovered in a real-world environment, particularly emergent and user-caused failures~\cite{gupta_failures_2017}. Testnets can reveal certain types of emergent and partial failures~\cite{fisman_formal_2022}. Relatively fewer novel types of failures can be revealed by experiments using emulation setups. In principle, simulations do not allow for discovering new types of failures.

\textbf{Reproducibility.} This parameter is tied to the availability of the same setup to different researchers. Simulation, at least in theory, allows for the highest level of reproducibility, given that all artefacts can easily be published. With emulations, the evaluated software can be made available, but access to an identical experimental testbed is not always available. It could be argued that while reproducibility is low for both testnets and deployed systems, a testnet allows for a relatively easier replication of experimental conditions, which is almost impossible with deployed systems.

\textbf{Control.} Simulation and emulation are the experimental setups that give researchers the most control over the flow of their experiments.
With testnets and real-world deployed systems, researchers usually have little to no control over the system while it is running since there is a dependency on the volunteers or end users that are running the software.

\textbf{Speed of change.} The speed at which an experiment can be modified is a distinguishing factor between different experimental setups. This speed of change is usually the lowest in a deployed real-world system, given the delays caused by the propagation of software updates. Testnets can allow for somewhat quicker deployment cycles. In both cases, deployment and the collection of results can be time-consuming. Simulation and emulation setups have relatively low external constraints and quickly be changed.

\textbf{Debugability.} Discovering, analyzing and reproducing software bugs in deployed systems require dedicated infrastructure and can be a time-consuming process. In comparison, testnets and emulation provide more debugability since the researcher has a higher level of control. Simulation can provide the highest discoverability of bugs as long as the scale of the simulation is not too large.

\begin{mytextbox}
\textit{This analysis shows that we can not have a comprehensive evaluation without deployment. We need to account for environmental realism and failures, which are detectable only in a real-world scenario.}
\end{mytextbox}

\begin{figure*}[t]
\centering
\includegraphics[width=.75\linewidth]{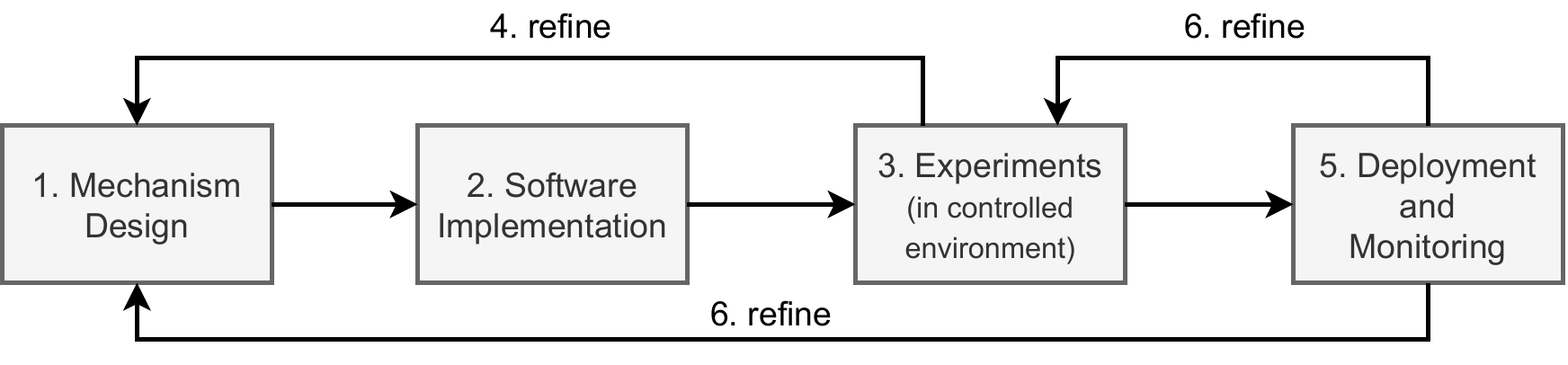}
\caption{Our deployment-first research methodology. The key difference with Figure~\ref{fig:traditional_methodology} is that we directly apply new insights or real-world workload traces to the mechanism design and experiments, respectively (step 6).}
\label{fig:our_methodology}
\end{figure*}

\section{Tribler: Deploying and Monitoring Decentralized Mechanisms at Scale}
\label{sec:tribler}
% We provide recommendations on what to do 

% OLD 
%TODO
%1) Introduce Tribler as experimental testbed and research vehicle -> learn-by-doing, learn-by-failing
%1a) Elaborate the importance of continuous monitoring and explain how we achieve this
%1b) Can we integrate our open-science approach in the storyline? Open Science is a necessity to adopt this methodology?!
%2) Bandwidth token economy as example

Tribler is our lab's peer-to-peer file-sharing software and research vehicle to deploy and evaluate decentralized mechanisms at scale~\cite{tribler_source_code}.
We have used the Tribler software for almost two decades to obtain unique insights into the complex interactions and dynamics in live peer-to-peer networks~\cite{pouwelse2008tribler}.
Over 30 PhD researchers and BSc/MSc students have used our software to evaluate their mechanisms.
Tribler also has a stable user base that enables longitudinal deployment experiments.
Over 1.8 million users have downloaded the Tribler software, and at the time of writing, Tribler has 40'000 unique monthly users.\footnote{See \url{https://release.tribler.org}}

Tribler was initially designed as a file-sharing application that allows users to download torrent files anonymously using a custom onion-routing protocol~\cite{jagerman2014fifteen}.
Tribler uses the IPv8 networking library that supports authenticated messaging and enables the construction and maintenance of decentralized overlays.
Over the years, however, Tribler has evolved from a BitTorrent download client to a versatile application with features such as keyword search, bundling torrents into channels, and reputation mechanisms to address free-riding behaviour~\cite{meulpolder2009bartercast}.

\subsection{Deployment-First Research Methodology}
Tribler is a vital part of our labs' research methodology since it enables us to deploy and evaluate decentralized mechanisms at scale.
Tribler also allowed us to try out a different research methodology with an increased focus on deployment.
Originating from our experiences with Tribler and deployment efforts, we now present our \emph{deployment-first} methodology of decentralized systems design and refinement based on a continuous experimentation cycle.
Figure~\ref{fig:our_methodology} visualizes an updated approach to the traditional research methodology shown in Figure~\ref{fig:traditional_methodology}.
We argue that in continuous experimentation, the system's deployment stage does not take place after experiments (step 3 in Figure~\ref{fig:our_methodology}).
Instead, we treat deployment as a critical next step in our research methodology that happens after experiments.
Potential findings from deployment studies include discovering new types of failures that do not occur in a controlled environment
and novel insights or performance issues caused by the unpredictability of real-world environments.
These insights feed directly into the refinement of the design and experiments in the following two ways (step 6 in Figure~\ref{fig:our_methodology}).
First, we leverage our new insights to update and improve the decentralized mechanism, similar to how the standard research methodology uses experimental results for refinement.
Second, we use information obtained from deployment to refine our experiments in a controlled environment.
This can be done, for example, by replaying a workload trace obtained from the live network during in-house experiments to evaluate mechanisms under a more realistic workload.
A key focus during deployment is on monitoring the mechanism to detect failure or anomalies.
This is further discussed in Section~\ref{sec:lessons_learned}.

Our methodology does not substitute the need for experiments in controlled environments.
On the opposite, data obtained from a deployment trial, such as network characteristics, the performance of clients, and user behaviour, should be used to address the limitations of other experimental setups, such as simulation and emulation.
Therefore, this data increases the realism of local experiments and helps in further validating mechanisms before deployment.%\todo{Maybe go through all table rows here briefly and show, how our methodology can address weaknesses and limitaion through combination of different experiments}. 

\subsection{Motivating Use-Case: Experimental Tokenomics}
\label{sec:use_case_tokenomics}

We now describe how we have applied our deployment-first methodology during a recent deployment trial.
This trial uses tokenomics to address free-riding behaviour while downloading content with Tribler.

\textbf{Mechanism Design and Objectives.}
A fundamental issue in peer-to-peer networks is free-riding behaviour, where one peer takes more resources from the community than it contributes~\cite{feldman2004free}.
In Tribler, this manifests as a user downloading more data from others than contributing back (seeding).
Earlier work established that free-riding behaviour in Tribler is typical, resulting in fewer uploaders and degradation of download speed~\cite{meulpolder2009bartercast}.
Since our anonymous downloading mechanism increases resource usage even further, addressing free-riding behaviour became an important issue as the Tribler network grew.

\begin{figure}[t]
\centering
\includegraphics[width=.95\linewidth]{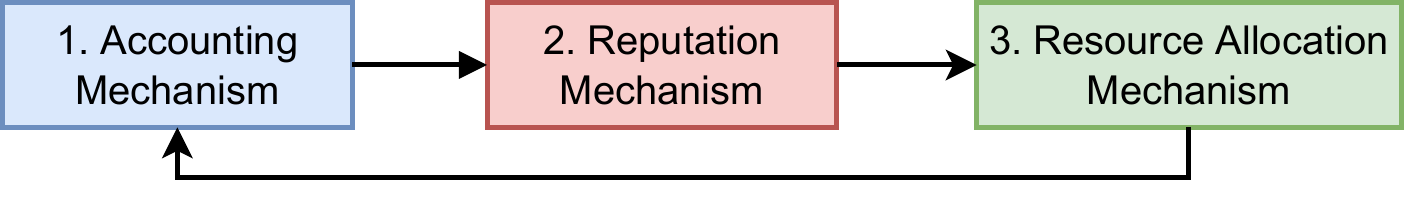}
\caption{Three complementary mechanisms we used to address free-riding behaviour in Tribler~\cite{de2021contrib}. We evaluated each mechanism using our deployment-first methodology.}
\label{fig:contrib_loop}
\end{figure}

Our solution to free-riding combines three complementary mechanisms, each designed, evaluated and deployed in Tribler using our deployment-first methodology (also see Figure~\ref{fig:contrib_loop}).
The first mechanism is a lightweight, decentralized ledger named TrustChain, which stores all pair-wise bandwidth transfers between users in the network in the form of records~\cite{otte2020trustchain}.
TrustChain is designed explicitly for lightweight accounting in decentralized networks and is highly scalable in the number of participants because it avoids a global consensus mechanism.
Users share these records with other users using a simple gossiping mechanism.
Our second mechanism is a reputation mechanism that, based on received records, computes a trustworthiness score for other users.
The third mechanism is a resource allocation mechanism that determines for each user to which other users it will upload data.
The combined working of these mechanisms allows users to identify free-riders themselves and consequentially refuse them services while giving honest users preferential treatment.
Additional details and experimental results can be found in our other work~\cite{de2021contrib}.

\textbf{Applying our Deployment-First Methodology.}
We deployed each of the three mechanisms and went through various deployment cycles to improve and fine-tune them.
As a first step, we designed TrustChain, implemented it and conducted correctness and validation experiments on our compute cluster (steps 1-3 in Figure~\ref{fig:our_methodology}).
We then integrated TrustChain into the Tribler software, implemented a crawler to gather created records, and published a new software release.
Due to the lack of real-world traces and insights, we could not adequately set some parameters, for example, the interval at which TrustChain records are shared with other users.
Only after a few deployment cycles did we have insights on setting such parameters.

We continuously monitored the created TrustChain records in the Tribler network, and we were able to detect various failures and design shortcomings that were not discovered during our local experiments.
For example, our deployment revealed that our initial design of TrustChain was falling short because a user can only be engaged in recording one transaction at the same time.
This shortcoming significantly limited the speed at which records could be created and is an essential limitation since the Tribler software frequently communicates with other users simultaneously.
It bootstrapped a redesign of the format of TrustChain records with support for concurrent transactions (see~\cite{de2021contrib}).
At the same time, we used the collected TrustChain records to start designing our reputation mechanism (see~\cite{otte2020trustchain}).
We also discovered various bugs in the deployment stage, for example, one bug was related to database corruption that occasionally occurs on a particular version of Windows.

%We describe three challenges we encountered while applying our methodology.

\subsection{Lessons Learned}
\label{sec:lessons_learned}
We have conducted multiple deployment trials with Tribler.
Due to space constraints, we cannot discuss all insights obtained when applying our deployment-first research methodology.
However, we will summarize four lessons we learned when working on the previously described use case and our other deployment trials.

%We learned that there need to be more ways to earn tokens. People were quite emotional over their token balances which we believe also indirectly affects how much they are downloading. \stef{how exactly did you learn that, does it count as a reliable user study?}

%Things that did not work:
%NOTE. FIVE LESSONS. STRUCTURE, REALISM CONDITION/FAILURES IDENTIFIED - INSIGHT. 

%- Wrong accounting, it did not reference real-world dynamics 
%- Was not secure 
%- With every update had to start all over again (challenge 1)
%- If it does not fail on its own, we need to try and break it, find the limitations of the system
%- We should use multiple metrics to see if the system is 'sane' (challenge 2)
%- challenge 3

%Interesting scenario: our collected datasets were not complete enough to always draw meaningful conclusions. Churn got in the way!

\textbf{Lesson I: Plan for Mechanism Upgrades and Maintaining Backwards Compatibility.}
Tribler consists of various mechanisms that we continuously monitor and improve.
Upgrading these mechanisms sometimes required us to make changes that break compatibility with prior versions.
This compatibility break results in the fragmentation of the network since users with different versions of a particular mechanism can no longer communicate with each other.
Additionally, such breaking changes often require software logic that updates locally stored data (e.g., in a database) to be compatible with the new mechanism.

We aim to minimize the number of breaking mechanism changes to avoid too much fragmentation of our network and to ensure sufficient usage of newly deployed mechanisms.
This aim is also motivated by our observation that users are relatively slow in updating their Tribler software when a new release is published, especially if the benefits of the software update are unclear.\footnote{See \url{https://release.tribler.org}.}
During our deployment trials, we learned that we should plan for mechanism upgrades already while designing a particular mechanism.
We note that this problem is not exclusive to Tribler since many blockchain systems occasionally have to upgrade their network protocol by releasing a new software version or forking the network, e.g., to fix security issues or improve performance~\cite{zamyatin2018wild}.

%The Tribler network is not only highly heterogeneous in terms of end-user compute power, but also in installed version of Tribler.
%The fragmentation of Tribler version leads to a few additional challenges.
%Firstly, we have to make sure that future version of Tribler remain backwards compatible with older Tribler versions.

%\textbf{Ensuring Robustness of Tribler.}
%Tribler is a complex application with several mechanisms that interact with each other.
%It is challenging to ensure that Tribler remains robust and that the number of crashes is minimized.
%To improve robustness, we use fuzzing techniques to expose our application to random inputs and to identify failures.
%However, failures can also be attributed to failing resources at the end-user side, for example, an unstable network connection or failing hard disks.

\textbf{Lesson II: The Importance of Monitoring.}
Continuously monitoring the behaviour of new mechanisms is critical to detect failures and anomalies in deployment~\cite{zeilemaker2011tribler}.
We engineer a crawler during every deployment trial and provision it when a new Tribler release is published.
This crawler joins a particular overlay network as a peer, sends data queries to other Tribler instances and persists the retrieved information in a local database.
This practice is comparable to collecting, analyzing and visualizing the transactions made in blockchain networks.

We have deployed multiple crawlers to gather data from our live network.
For example, alongside the TrustChain ledger, we also deployed a crawler that collects records created by users.
However, due to churn, the crawler sometimes is unable to collect particular data points.
Because a user could have gone offline before the crawler sent a request, our datasets did not always contain all the data points we required.
Despite this, the data collected during deployment revealed a large-scale outage due to a software bug since the number of created TrustChain records dropped significantly.
These experiences taught us that monitoring infrastructure is crucial to planning a deployment.

%\stef{aren't these all reasons against deployment because your data is not reliable?}

%A key objective of Tribler is to be a download client with privacy-enhancing features.
%As such, we should be particularly careful during the monitoring stage of our algorithms that we collect only the sufficient information required to obtain scientific insights and improve our algorithms.
%When not possible, we add noise to collected data to prevent the identification and behavior patterns of individual users.

\textbf{Lesson III: Document all Design Decisions and Changes.}
Successfully applying our deployment-first methodology requires adequate planning and introduces unique challenges for developers and researchers.
In early deployment trials, we could have documented our design and deployment decisions better and, therefore, would have avoided repeating prior mistakes.
Over the years, we adopted the \emph{open science} approach~\cite{cruwell2019seven} to publicly record all our source code, design decisions and meeting minutes.
We also carefully report our observations from the deployment environment and document failures to avoid repeating particular mistakes in future iterations of a mechanism.
This open science approach is now an essential aspect of our Tribler development cycle and research methodology.
Open science also helps other researchers understand and replicate our prior results. It is also beneficial for users interested in understanding how the Tribler software behaves, what data is being collected, and what mechanisms are being executed on their devices.
All this information is publicly available on our GitHub repository.\footnote{See \url{https://github.com/tribler/tribler/issues}.}
%We also note that this form of documentation is increasingly being used by blockchain research labs.\stef{evidence?}

\textbf{Lesson IV: Do not Deploy Too Much at Once.}
A common mistake we made during early deployment trials was that we tended to include multiple new features or mechanisms in a single release.
Not only did this prolong the time between software releases, but it also increased the risk of breaking the Tribler software when there was a defect in one of the newly-deployed mechanisms.
It also made it impossible to isolate the effects of specific changes. 
To avoid these risks, we currently aim to include at most one new feature per release and aim for short release cycles.
For example, we shipped each of the mechanisms described in the use case in Section~\ref{sec:use_case_tokenomics} with separate releases with a few months between them.

We also learned that mechanism design is an incremental process that requires multiple iterations to grow and become fruitful.
For example, when designing a socio-economic mechanism, it is often impossible to adequately parameterize the mechanism since the dynamics of the deployment environment are not known apriori by the researcher.
Only in response to data collected from a real-world environment the mechanism can be made robust and optimized for a particular application domain.
%Even though one can rely on game-theoretical models to reason about and predict the behavior of users, rationality of users cannot always be assumed.
%For example, faulty software might violate some assumptions made by the practitioner.\todo{think about this}

% - Real-world dynamics are hard to predict, rationality by end-users cannot always be assumed and as such, game-theoretic models cannot always be applied.
 
% Challenges encountered when applying our deployment-driven methodology:
% - Dealing with user fragmentation, users with different users have to interact with each other.
% - Consequentially, Tribler consists of various algorithms that might be interconnected. Failure in one can lead to failure in other components.
% - Dealing with upgrades to the protocol which sometimes requires a clean-slate approach.
% - Security -> we made the decision to

%\textbf{Lessons Learned.}
%We learned that there need to be more ways to earn tokens. People were quite emotional over their token balances which we believe also indirectly affects how much they are downloading. \stef{how exactly did you learn that, does it count as a reliable user study?}

%Things that did not work:
%- Wrong accounting, it did not reference real-world dynamics 
%- Was not secure 
%- With every update had to start all over again
%- If it does not fail on its own, we need to try and break it, find the limitations of the system
%- We should use multiple metrics to see if the system is 'sane'

%Interesting scenario: our collected datasets were not complete enough to always draw meaningful conclusions. Churn got in the way!

\section{The Road Ahead}

We have argued that the increasing complexity and dependencies on decentralized systems such as blockchain applications require more robust and mature experimentation methodologies. Such methodologies are needed to identify new types of failures in realistic environments. We argued that deployment should be explicitly integrated as a key step in the research methodology to improve the evaluation of decentralized mechanisms.

We have presented our \emph{deployment-first approach} that goes beyond the standard research methodologies. 
We also presented Tribler, our research vehicle for deploying decentralized mechanisms.
We showed how we use insights from deployment trials to improve the design of decentralized mechanisms and their experiments.
We have shown that experimental setups based on deployment provide (1) insights into new types of failures; and (2) a foundation for the design of realistic experiments in controlled environments.
By describing a tokenomics use case, we demonstrated the feasibility of our deployment-first approach in practice.

Our deployment-first approach is a continuously evolving methodology. One possible extension is the addition of infrastructure and approaches for A/B testing decentralized mechanisms. This approach would serve different algorithms and parameters to distinct subsets of users.
%such as blockchain fabrics and consensus mechanisms, we argue that there should be increased research effort to devise platforms to quickly deploy decentralized mechanisms.
%Building unified infrastructure for the deployment of such mechanisms, like Tribler, has initial development costs associated with it but these costs would diminish over time as the platform matures.
%Having an open deployment infrastructure makes our deployment-first methodology accessible to more research labs.

%One possible extension to speed up our deployment-first approach would be a solid infrastructure for A/B testing of decentralized mechanisms, where algorithms with different parameters would be served to distinct subsets of users.

%CONCLUDING ARGUMENTS

%There are more and more deployments but not enough rigorous experimentation and methodology. We provide this methodology. We show that this is possible in practice. We show that limits of different experimental setups can be mitigated (using table). 

%In decentralized complex systems.  

%We show that deployment is necessary because of failure detection and realism (you can not even bootstraps controlled env without real world data, you do not always have data sets)

%there is no similar enough application domain from which to harvest data)

%Takeaway message: Failure in systems should be systematically dosumented and treated as deployment-stage experiments 

%Cornerstone of our lab research methodology. We leverage the best of both worlds. We acknowledge it is more work, but it is worth it. 

\bibliography{references}{}

\begin{thebibliography}{10}

\bibitem{berger2022does}
Christian Berger et~al.
\newblock Does my bft protocol implementation scale?
\newblock In {\em DICG}, 2022.

\bibitem{blackburn_cooperation_2022}
Alyssa Blackburn et~al.
\newblock Cooperation among an anonymous group protected {Bitcoin} during
  failures of decentralization.
\newblock {\em arXiv}, 2022.

\bibitem{blair_complex_2018}
Gordon Blair.
\newblock Complex {Distributed} {Systems}: {The} {Need} for {Fresh}
  {Perspectives}.
\newblock In {\em ICDCS}, 2018.

\bibitem{buchert_survey_2015}
Tomasz Buchert et~al.
\newblock A survey of general-purpose experiment management tools for
  distributed systems.
\newblock {\em FGCS}, 2015.

\bibitem{fisman_formal_2022}
Franck Cassez et~al.
\newblock Formal verification of the ethereum 2.0 beacon chain.
\newblock In {\em TACAS'22}, 2022.

\bibitem{cruwell2019seven}
Sophia Cr{\"u}well et~al.
\newblock Seven easy steps to open science.
\newblock {\em Zeitschrift f{\"u}r Psychologie}, 2019.

\bibitem{de2021contrib}
Martijn de~Vos et~al.
\newblock Contrib: Maintaining fairness in decentralized big tech alternatives
  by accounting work.
\newblock {\em Computer Networks}, 2021.

\bibitem{tribler_source_code}
Martijn de~Vos et~al.
\newblock Tribler/tribler: v7.12.1.
\newblock \url{https://doi.org/10.5281/zenodo.7097138}, September 2022.

\bibitem{feldman2004free}
Michal Feldman et~al.
\newblock Free-riding and whitewashing in peer-to-peer systems.
\newblock In {\em PINS workshop}, 2004.

\bibitem{ethereumgrants}
Ethereum Foundation.
\newblock Announcing ethereum foundation and co-funded grants.
\newblock
  \url{https://blog.ethereum.org/2019/08/26/announcing-ethereum-foundation-and-co-funded-grants},
  2019.
\newblock Accessed: 2023-01-11.

\bibitem{Gabriel_2006}
Richard~P. Gabriel et~al.
\newblock Ultra-large-scale systems.
\newblock In {\em OOPSLA}, 2006.

\bibitem{garraghan_emergent_2018}
Peter Garraghan et~al.
\newblock Emergent {Failures}: {Rethinking} {Cloud} {Reliability} at {Scale}.
\newblock {\em Cloud Computing}, 2018.

\bibitem{gupta_failures_2017}
Saurabh Gupta et~al.
\newblock Failures in large scale systems: long-term measurement, analysis, and
  implications.
\newblock In {\em Supercomputing}, 2017.

\bibitem{gustedt_experimental_2009}
Jens Gustedt et~al.
\newblock Experimental methodologies for large-scale systems: a survey.
\newblock {\em Parallel Processing Letters}, 2009.

\bibitem{Haeberlen2006FallaciesIE}
Andreas Haeberlen et~al.
\newblock Fallacies in evaluating decentralized systems.
\newblock In {\em International Workshop on Peer-to-Peer Systems}, 2006.

\bibitem{jagerman2014fifteen}
Rolf Jagerman et~al.
\newblock The fifteen year struggle of decentralizing privacy-enhancing
  technology.
\newblock {\em arXiv}, 2014.

\bibitem{javadi_failure_2013}
Bahman Javadi et~al.
\newblock The {Failure} {Trace} {Archive}: {Enabling} the comparison of failure
  measurements and models of distributed systems.
\newblock {\em JPDC}, 2013.

\bibitem{marijan_blockchain_2022}
Dusica Marijan and Chhagan Lal.
\newblock Blockchain verification and validation: Techniques, challenges, and
  research directions.
\newblock {\em Computer Science Review}, 2022.

\bibitem{meulpolder2009bartercast}
Michel Meulpolder et~al.
\newblock Bartercast: A practical approach to prevent lazy freeriding in p2p
  networks.
\newblock In {\em ISPDP}, 2009.

\bibitem{nasrulin2022gromit}
Bulat Nasrulin et~al.
\newblock Gromit: Benchmarking the performance and scalability of blockchain
  systems.
\newblock In {\em DAPPS}, 2022.

\bibitem{otte2020trustchain}
Pim Otte et~al.
\newblock Trustchain: A sybil-resistant scalable blockchain.
\newblock {\em FGCS}, 2020.

\bibitem{pouwelse2008tribler}
Johan Pouwelse et~al.
\newblock Tribler: a social-based peer-to-peer system.
\newblock {\em Concurrency and computation: Practice and experience}, 2008.

\bibitem{sarzyniec2013design}
Luc Sarzyniec et~al.
\newblock Design and evaluation of a virtual experimental environment for
  distributed systems.
\newblock In {\em PDP}, 2013.

\bibitem{tessone_stochastic_2021}
Claudio~J Tessone et~al.
\newblock Stochastic modelling of blockchain consensus.
\newblock {\em arXiv}, 2021.

\bibitem{vespignani_predicting_2009}
Alessandro Vespignani.
\newblock Predicting the {Behavior} of {Techno}-{Social} {Systems}.
\newblock {\em Science}, 2009.

\bibitem{wang_what_2017}
Guosai Wang et~al.
\newblock What can we learn from four years of data center hardware failures?
\newblock In {\em DSN}, 2017.

\bibitem{zamyatin2018wild}
Alexei Zamyatin et~al.
\newblock A wild velvet fork appears! inclusive blockchain protocol changes in
  practice.
\newblock In {\em Financial Cryptography}, 2018.

\bibitem{zeilemaker2011tribler}
Niels Zeilemaker et~al.
\newblock Tribler: P2p media search and sharing.
\newblock In {\em ACM Multimedia}, 2011.

\bibitem{zhou_sok_2022}
Liyi Zhou et~al.
\newblock Sok: Decentralized finance (defi) attacks.
\newblock {\em Cryptology ePrint Archive}, 2022.

\end{thebibliography}
\bibliographystyle{plain}

\end{document}